\documentclass{ws-mplb}

\begin{document}

\markboth{B.J. Ahmedov} {On a Possibility to Measure
Thermoelectric Power in SNS Structures}

\catchline{}{}{}{}{}

\title{ON A POSSIBILITY TO MEASURE THERMOELECTRIC POWER IN SNS STRUCTURES}

\author{\footnotesize B.J. AHMEDOV}

\address{Institute of Nuclear Physics and Ulugh Beg Astronomical Institute\\ Ulughbek,
Tashkent 100214, Uzbekistan\\
ahmedov@astrin.uzsci.net}

\maketitle


\begin{abstract}
Two dissimilar Josephson junctions, which are connected to a
heater can act as precise batteries. Because of the difference in
thermoelectric power of these batteries, circuit with two
dissimilar batteries, under heat flow $\Delta T\sim 10^{-5}K$
would have a net EMF $10^{-11} V$ around the zero-resistance loop
leading to a loop's magnetic flux oscillating in time. It is shown
its theoretical value is proportional to both the temperature
difference as well as the disparity in the thermoelectric powers
of the two junctions.

\end{abstract}

\ccode{PACS Number(s).: 74.50+r, 42.50.Dv}


\vspace*{1cm}

Thermoelectric effects in superconductor - normal metal -
superconductor ($SNS$) junctions have been studied theoretically
and observed experimentally in the papers$^{1-5}$. The purpose of
this short Letter is to suggest from theoretical considerations a
new proposal to measure thermoelectric effects in the SNS
structures.

We shall first consider the behaviour of a single Josephson
junction when temperatures of its electrodes $S$ have difference $\Delta T$
or, in other words, when there is a heat flow through the junction.
We use
resistively shunted model of the Josephson junction in which the latter is
considered as a circuit made up of the superconducting junction itself and a
normal resistance connected in parallel. Then, due to the Kirchoff's first
law,  $\vec\jmath _{s}=\vec\jmath _{n}$.

 According to Ohm's law a normal component of a current
in the junction is
$\vec\jmath _{n}=\lambda \vec E-\lambda grad\mu_e + \lambda\beta grad T$,
where $\lambda$ is electrical conductivity, $\mu_e$ is the chemical
potential per unit charge and $\beta$ is thermoelectric power.

Density of the superconducting current flowing through the junction is
related to the phase difference $\phi=\Delta\vartheta$ of
superconducting wavefunction  $\psi=n_s^{1/2}e^{i\vartheta}$
across the junction by$^6$
$\vec\jmath _{s}=\vec\jmath _{c}\sin\phi$,
where $\vec\jmath_{c}$ is the critical value of electric
current density, $n_s$ and $m_s$ are the density and mass
of Cooper pairs.

Using the considered formulae and Josephson equation$^6$
\begin{eqnarray}
-\hbar\frac{\partial\phi}{c\partial t}=\frac{2e}{c}\mu_e-\frac{2e}{c}\varphi
\end{eqnarray}
one can obtain
\begin{eqnarray}
\vec\jmath_{c}\sin\phi =\lambda grad
(\frac{\hbar}{2e}\frac{\partial\phi}{\partial t})+
\lambda \beta grad T\nonumber
\end{eqnarray}
which after integration on $\vec n dS$ will give
\begin{eqnarray}
RI_{c}\sin\phi
=\frac{\hbar}{2e}\frac{\partial\phi}{\partial t}+
\beta\Delta T.
\end{eqnarray}
Here $R=\int{dl}/{\lambda dS}$ is the resistance of the normal
layer with length $dl$, $\varphi$ is the scalar potential and
$\vec A$ is the vector-potential of electromagnetic field,
$I_{c}=\int \vec\jmath_{c}\vec n dS$ is electric current, $\vec n$
is normal vector to the cross section of wire $dS$.

Therefore if thermoelectric current exceeds the critical
current of the Josephson junction, then as a consequence of (2),
an alternating current (ac) of frequency
\begin{eqnarray}
\omega=\frac{2e}{\hbar}\beta\Delta T
\end{eqnarray}
is produced.
Formula (3) describes thermoelectric ac
Josephson effect according to which a temperature difference $\Delta T$
across the $SNS$ junction results in the Josephson oscillation with the
frequency $\omega$, which has
been predicted in $^{2}$ and experimentally confirmed $^{3}$.

Consider now two dissimilar $SNS$ junctions $I$ and $II$, which are
connected in opposition by superconducting wires to a common heater source
and form a loop with inductance $L$. Josephson junctions have unequal
thermoelectric powers $\beta_I$ and $\beta_{II}$, respectively,
since they are made from the different materials.
According to two-fluid model two currents flow in the superconductor: the
superconducting current of density $\vec\jmath_s$ and the normal current
of density $\vec\jmath_n$.
In the presence of heat flow, the total current density in the interior of
a bulk isotropic homogeneous superconductor must be zero (see,
for review$^{7-10}$) $\vec\jmath=\vec\jmath_s+\vec\jmath_n=0$ and so
\begin{eqnarray}
\vec\jmath_n=\lambda\beta gradT=-\vec\jmath_s=\frac{2n_se}{m_s}[\hbar
grad\vartheta-\frac{2e}{c}\vec A].
\end{eqnarray}

Now we integrate equation (4) over the contour $l$ which passes through
the interior of superconducting
ring with two Josephson contacts and use that $\oint\vec Ad\vec l=\int\vec
Bd\vec S=\Phi_b$ and $\oint grad\vartheta d\vec l=2\pi n$, where
$n$ is integer. Then
\begin{eqnarray}
\Phi_b=n\Phi_0+\frac{\hbar c}{2e}(\phi_{II}-\phi_{I}),
\end{eqnarray}
where $\phi_{II}$ and $\phi_{I}$ are the contributions due to the phase
discontinuities ($\phi=\Delta\vartheta$) at the Josephson junctions
 $II$ and $I$,
$\Phi_0=\pi\hbar c/e=2\times 10^{-7} Gauss\cdot cm^2$ is quantum of the
magnetic flux.

The rate of change of magnetic flux is related to $\Delta\omega
\equiv \omega_{II}-\omega_I$ by
\begin{eqnarray}
\frac{d\Phi_b}{dt}=\frac{dn}{dt}\Phi_0+\frac{\hbar c}{2e}\Delta\omega ,
\end{eqnarray}
so if $\Delta\omega$ is not equal to zero, a magnetic flux
$\Delta\Phi_b\ne 0$ will be induced. As long as $\Delta\Phi_b<\Phi_0$,
$n$ will remain constant and $\Delta\Phi_b$ will increase linearly with
time until $\Delta\Phi_b=\Phi_0$, then the order of the step $n$ will
change as flux quantum enters the loop. Thus due to the effect of
gradient of temperature on the junctions with unequal
thermoelectric powers is equivalent to having a time dependent flux, as
given by the last term on the left hand side of equation (6).
Using equations (3) and (6) one can derive that the change in
magnetic flux $\Phi_b$ inside the circuit during the time interval $[0,t]$
subjects to formula
\begin{eqnarray}
\Delta\Phi_b=\Phi_0\Delta n+c\int_0^t\Delta T(\beta_{II}-\beta_{I})dt.
\end{eqnarray}

Thus this particular loop is sensitive to the frequency and in this
connection to the thermoelectric power difference between the junctions.

When the current exceeds the critical value potential
difference $V=\beta\Delta T$ appears across the junction due to the
thermoelectric effects.
Since the thermoelectric power of the junction $II$ differs from that at
the junction $I$, the potential differences across the first and second
junctions, $V_{II}$ and $V_I$, respectively, will differ so that
$\Delta V=V_{II}-V_I=(\beta_{II}-\beta_{I})\Delta T.$ The basic technique
for the detection of extremely small voltage differences between two
Josephson junctions by monitoring of magnetic flux change was firstly
developed by Clarke$^{12}$.

In the absence of any additional effects on the Cooper pairs, one
would thus expect the net EMF in the loop containing the junctions
to be $(\beta_{II}-\beta_{I})\Delta T\sim 10^{-11} V$ for the
typical values of parameters $(\beta_{II}-\beta_{I})\sim 10^{-6}
V/K$ and $\Delta T\sim 10^{-5}K$. For the loop of inductance $L$
the evolution of magnetic field is approximately governed by law
${d\Phi_b}/{dt}=-L{dI_l}/{dt}$. In this connection a nonvanishing
value for $\Delta V$ would lead, according to (3) and (6), to a
time varying current $I_l$ (from zero to the critical maximum
value in the range of one number of the step $n$):
${dI_l}/{cdt}=-{1}/{L}\Delta V$, which will induce the above
discussed flux $\Delta\Phi_b=c\int\Delta Vdt$ through the loop in
the linear regime.

There are at least two realistic processes which change the ideal behaviour
magnetic field $\Phi_b$ described by equation (7): (i) the ohmic
dissipation of current in the normal layer of the Josephson contacts
and (ii) the dependence of maximum superconducting current, which can
pass through the junctions without dissipation on the magnetic field.

A new experiment, in which the thermoelectric response creates
saw-tooth flux (7) oscillating in time can be proposed. It will
give a good opportunity to measure thermal effects in SNS
structures and may be relevant for confirmation of some aspects of
thermoelectric transport theory in the superconducting state. Let
us roughly evaluate the possibilty of performing this experiment.
Taking $(\beta_I-\beta_{II})\sim 10^{-8}cm^{1/2}\cdot g^{1/2}\cdot
s^{-1}/K$ and $\Delta T\sim 10^{-5}K$ we obtain for
$\Delta\Phi_b(Gauss\cdot cm^2)\sim 3\times 10^{-3} \cdot \Delta t
(s)$ in the range of one step $n$. Measuring such variations of
magnetic flux is in the capacity of SQUID sensitivity and can be
easily performed. The obtained result, in principle, may be used
for construction of new type devices such as high sensitivity
precise thermometers and thermocouples.

The main problem in observation in the flux (7) will be connected with
generating thermal current $\vec\jmath_s=-\lambda\beta
gradT$ comparable in magnitude with its critical value $\vec\jmath_c$,
since the temperature difference across the junction is limited by a low
temperature $T_c$ and small sizes of the junction.

The similar method of measurement has been used by Jain et al$^{12}$ in null
result experiment on
confirmation of the strong equivalence principle for a charged
massive particle. In their experiment the
phase of Josephson contacts has been locked to an external  microwave source.

It is interesting to mention that the predicted mechanism for production
of magnetic field and current changing with time can be of crucial importance
in astrophysics as an additional way (to the existed ones$^{13}$) for
generation of electromagnetic radiation from pulsars. According to the
recent theoretical models$^{14}$, a neutron star is the relativistic compact
object consisting of the conducting crust and superfluid core. In the
inner crust of the neutron star the superfluid coexists with a crystal
lattice and in its core, at densities above $2\times 10^{14} gm/cm^3$
there is a homogeneous mixture of superfluid neutrons and superconducting
protons.

Important fact is that the thermoelectric power $\beta$ is the
function of temperature as $T^{3/2}$ and in this connection can
reach large numbers since superconductivity in the stars takes
place at the temperatures $10^6\div 10^{7}K$. So if one accepts
that the $SNS$ structures are realized in the intermediate
boundary between conducting crust and superconducting core inside
the neutron star then the strong heat fluxes in these $SNS$
junctions can lead to the generation of time-varying magnetic
field (i.e. electromagnetic radiation) due to the thermoelectric
effect described by the basic formula (7).

\section*{Acknowledgments}

The author acknowledges the financial support and hospitality at the
Abdus Salam International Centre for Theoretical Physics, Trieste, where
the main part of the work was done.

\begin{flushleft}

{\bf References}\\

1. J. Clarke and S.M. Freake, {\it Phys. Rev. Lett. \bf 29}, 588 (1972).\\

2. A.G. Aronov and Yu.M. Galperin, {\it JETP Lett. \bf 19}, 165 (1974).\\

3. G.I. Panaitov, V.V. Ryazanov, A.V. Ustinov and V.V. Schmidt,
     {\it Phys. Lett.A \bf 100}, 301 (1984).\\

4. V.V. Shmidt, {\it JETP Lett. \bf 33}, 98 (1981).\\

5. V.V. Ryazanov
and V.V. Schmidt, {\it Solid State Comm. \bf 40}, 1055 (1981).\\

6. B.D. Josephson, {\it Phys. Lett. \bf 1}, 251 (1962).\\

7. V.L. Ginzburg and G.F. Zharkov, {\it Sov. Phys. Uspekhi \bf 21},
381 (1978).\\

8. D.J. Van Harlingen, {\it Physica \bf 109\&110B}, 1710 (1982).\\

9. R.P. Huebener, {\it Supercond. Sci. Technol. \bf 8}, 189 (1995).\\

10. Z.D. Wang, Q. Wang and P.C.W. Fung, {\it Supercond. Sci. Tecnol. \bf 9},
333 (1996).\\

11. J. Clarke, {\it Phys. Rev. Lett. \bf 21}, 1566 (1968).\\

12. A.K. Jain, J. Lukens J and J.S. Tsai, {\it Phys. Rev. Lett. \bf 58},
1165 (1987).\\

13. F.C. Michel, {\it Theory of Neutron Star Magnetospheres} (Univ.
Chicago Press, 1991).\\

14. S.L. Shapiro  and S.A. Teukolsky, {\it Black Holes, White
Dwarfs, and
Neutron Stars} (New York: Wiley, 1983).\\

\end{flushleft}

\end{document}